\documentclass[prd,amssymb,amsmath,amsfonts,superscriptaddress,nofootinbib,reprint,showpacs]{revtex4-1}

\usepackage{graphicx}
\usepackage{lmodern}
\usepackage{amsmath,amssymb}
\usepackage{mathrsfs}
\usepackage{amsfonts}
\usepackage[utf8]{inputenc}
\usepackage{url}
\usepackage[colorlinks]{hyperref}
\usepackage{xcolor}

\newcommand{\Birmingham}{School of Physics and Astronomy and Institute for Gravitational Wave Astronomy, University of Birmingham, Edgbaston, Birmingham, B15 9TT, United Kingdom}
\newcommand{\GRAPPA}{\affiliation{GRAPPA, Anton Pannekoek Institute for Astronomy and Institute of High-Energy Physics, University of Amsterdam, Science Park 904, 1098 XH Amsterdam, The Netherlands}}
\newcommand{\DeltaITP}{\affiliation{Delta Institute for Theoretical Physics, Science Park 904, 1090 GL Amsterdam, The Netherlands}}
\newcommand{\fm}{\texttt{fmtidal}}

\begin{document}
%%%%%%%%%%

\title{Frequency domain model of $f$-mode dynamic tides in gravitational waveforms from compact binary inspirals}

\author{Patricia Schmidt}
\email{pschmidt@star.sr.bham.ac.uk}
\affiliation{\Birmingham}

\author{Tanja Hinderer}
\email{t.hinderer@uva.nl}
\GRAPPA\DeltaITP

%%%%%%%%%%%%%%%
\begin{abstract}
The recent detection of gravitational waves (GWs) from the neutron star binary inspiral GW170817 has opened a unique avenue to probe matter and fundamental interactions in previously unexplored regimes. Extracting information on neutron star matter from the observed GWs requires robust and computationally efficient theoretical waveform models. We develop an approximate frequency-domain GW phase model of a main GW signature of matter: dynamic tides associated with the neutron stars' fundamental oscillation modes ($f$-modes). We focus on nonspinning objects on circular orbits and demonstrate that, despite its mathematical simplicity, the new ``$f$-mode tidal'' (\fm) model is in good agreement with the effective-one-body dynamical tides model up to GW frequencies of $\gtrsim 1$ kHz and gives physical meaning to part of the phenomenology captured in tidal models tuned to numerical relativity. The advantages of the \fm\,model are that it makes explicit the dependence of the GW phasing on the characteristic equation-of-state parameters, i.e., tidal deformabilities \emph{and} $f$-mode frequencies; it is computationally efficient; and it can readily be added to any frequency-domain baseline waveform. The \fm\,model is easily amenable to future improvements and provides the means for a first step towards independently measuring additional fundamental properties of neutron star matter beyond the tidal deformability as well as performing novel tests of general relativity from GW observations. 
\end{abstract}

%%%%%%%%%%%%%%%
\maketitle 
%%%%%%%%%%%%%%%
\section{Introduction}
\label{sec:intro}
%%%%%%%%%%%%%%%
The first observation of gravitational waves (GWs) from the inspiraling neutron star (NS) binary GW170817~\cite{GW170817} initiated using GWs to elucidate long-standing questions in subatomic physics~\cite{Lattimer:2004pg,Potekhin:2011xe,Baym:2017whm}. This event enabled constraining, for the first time, the equation of state (EoS) of NS matter from tidal effects in the GW signal~\cite{Abbott:2018wiz,Abbott:2018exr}. Extracting the information on the fundamental properties of NS matter from the GW data requires robust theoretical waveform models that are accurate over a wide range of parameters, computationally efficient, and include all relevant physical effects.  

The presence of matter gives rise to a number of different GW signatures compared to signals from black hole (BH) binaries (see e.g.~\cite{Rezzolla:2018jee,Barack:2018yly}). Here, we focus on a subset of tidal effects during a binary inspiral that are associated with the response of matter to the spacetime curvature sourced by the companion. Specifically, we consider the GW signature from the tidal excitation of the objects' fundamental oscillation modes ($f$-modes), which is characterized by two parameters for each $\ell-$th multipolar mode: the tidal deformability $\lambda_\ell$~\cite{Flanagan:2007ix,Hinderer:2007mb,Binnington:2009bb,Damour:2009vw} and the angular fundamental-mode frequency\footnote{Oscillation modes are characterized by three integers $(n,\ell,m)$, where $n$ denotes the number of radial nodes in the mode function and $m$ is the azimuthal integer. In the nonspinning case, the mode frequency is independent of $m$, and for the $f$-modes $n=0$, hence our notation $\omega_{n\ell m}\mid_{ f\mathrm{-mode}}=\omega_{\ell}$.} $\omega_{\ell}$~\cite{Kokkotas:1999bd}. In general relativity and for a range of proposed EoS models, the parameters $\lambda_\ell$ and $\omega_{\ell}$ are related by approximately universal relations (UR)~\cite{Chan:2014kua}. Measuring both parameters simultaneously can thus provide important insights into the fundamental properties of matter and represents a first step towards GW asteroseismology of NSs, where potential future measurements of tidally excited NS oscillation modes could enable discerning details of the complex physics of their interiors~\cite{Andersson:1997rn}.

Tidal effects associated with the $f$-modes have previously been included in gravitational waveform models in different ways. Analytical models have primarily focused on adiabatic tidal effects~\cite{Flanagan:2007ix,Damour:2009wj, Bini:2012gu, Vines:2011ud, Damour:2012yf, Bini:2014zxa, Bernuzzi:2014owa, Nagar:2018zoe} describing the regime where $\omega_\ell$ is much higher than any other frequency in the system. Finite-$\omega_\ell$ effects, known as dynamical tides, become most important in the late inspiral, when $\ell$-th multiples of the orbital frequency characterizing the tidal forcing frequency become comparable to the $f$-mode frequency, thus approaching resonance. Although the resonance itself is often at relatively high frequencies ($\gtrsim 1$ kHz), the $f$-mode signatures in the GW phase start to accumulate long before the resonance~\cite{Hinderer:2016eia,Steinhoff:2016rfi, Flanagan:2007ix,Hinderer:2009ca}.
These effects have been considered in Newtonian gravity~\cite{Shibata:1993qc,Ho:1998hq, Flanagan:2007ix,Lai:1993di,Kokkotas:1995xe,1994ApJ...426..688R} and within the time-domain tidal effective-one-body (EOB) model of Refs.~\cite{Hinderer:2016eia, Steinhoff:2016rfi}, hereafter TEOB. EOB models require solving for the time evolution of the binary inspiral to obtain the GW signal, which is computationally expensive and makes the dependence on parameters less transparent. More efficient waveform models are those that directly provide a description of the GWs in the frequency domain, either from phenomenological models~\cite{Ajith:2007qp,Ajith:2007kx,Ajith:2009bn,Santamaria:2010yb,Hannam:2013oca,Khan:2015jqa}, reduced-order models for EOB~\cite{Field:2013cfa, Purrer:2015tud, Bohe:2016gbl, Lackey:2016krb}, or surrogates of NR waveforms~\cite{Blackman:2017dfb,Blackman:2017pcm}. 
Frequency-domain models calibrated to numerical-relativity (NR) simulations in the tidal sector~\cite{Lackey:2016krb,Dietrich:2017aum,Kawaguchi:2018gvj,Dietrich:2018uni} phenomenologically describe an enhancement of matter effects compared to predictions from adiabatic tidal models; however, these models depend only on the tidal deformability parameters and are restricted in their parameter space coverage, and hence are of limited applicability. 

In this paper we aim to advance the physics content of frequency-domain tidal models and thus enhance the scope of science that can be done with GW observations by developing an approximate analytic model of Newtonian dynamical $f$-mode tides for efficient GW data analysis. We demonstrate that for a nonspinning quasicircular inspiral the dynamical $f$-mode effects can be captured by a simple, closed-form expression that adds linearly to the known adiabatic tidal effects in the phase~\cite{Flanagan:2007ix, Vines:2011ud, Damour:2012yf}. We show that our simple model, which we refer to as ``$f$-mode tidal'' (\fm), agrees well with the TEOB model for a range of EoSs. We further compare to two NR-calibrated models~\cite{Dietrich:2017aum,Kawaguchi:2018gvj,Dietrich:2018uni} and find agreement with~\cite{Kawaguchi:2018gvj} over most of the inspiral, indicating that part of the tidal enhancement over the adiabatic models that is captured phenomenologically by NR-calibrated terms is consistent with dynamical $f$-mode tides. 
This is important for determining the choice of parameters in GW models and is expected to enhance the robustness of efficient models over a wider range in parameter space, potentially helping us to mitigate systematic errors that will become significant as GW detectors improve in sensitivity~\cite{Samajdar:2018dcx, Abbott:2018wiz}. Moreover, accounting for more realistic physics such as the $f$-mode will render efficient GW models applicable to more general contexts and enable novel tests of the fundamental physics of NSs, exotic objects, and alternative theories of gravity.

The \fm\,model derived in this paper can be directly included in \emph{any} frequency domain point-particle baseline model for a binary inspiral. Other matter effects such as spin-induced multipoles, gravitomagnetic tides, spin-tidal interactions, and tidal heating are already known and enter separately into these models~\cite{Jimenez-Forteza:2018buh,Banihashemi:2018xfb,Krishnendu:2017shb,Poisson:1997ha,Isoyama:2017tbp,Abdelsalhin:2018reg,Landry:2018bil}. The \fm\, model can easily be improved with inputs from NR for the behavior at higher frequencies, where other physics may become important. We emphasize that \fm\,is specialized to nonspinning binaries on circular orbits where the $f$-mode resonance occurs near the end of the inspiral. 
We leave to future work the inclusion of  relativistic corrections~\cite{Steinhoff:2016rfi,Steinhoffinprep} and spin effects on the $f$-mode~\cite{Ho:1998hq,Doneva:2013zqa,Steinhoffinprep}. A Newtonian estimate~\cite{Ho:1998hq} indicates that spin-induced shifts of the $f$-mode resonance could change the orbital phase by $\sim 10 -1000$ cycles in extreme cases. Relativistic corrections are expected to lead to a net enhancement of the $f$-mode effect and include frame-dragging and redshift-induced effective shifts of the resonance, an increased tidal field, and reduced GW dissipation; once theoretical results~\cite{Steinhoffinprep} for these corrections are available they can be used to refine the \fm\, model. Future work will also need to consider passage through resonance that is required for generalizing to other NS modes beyond the $f$-modes~\cite{Nollert:1999ji} and to eccentric orbits which lead to a striking $f$-mode effect~\cite{Gold:2011df,Chirenti:2016xys,Yang:2018bzx}.

We note work on a similar topic, containing complementary methods and results, recently appeared~\cite{Andersson:2019dwg}.

Throughout this paper we set $G=c=1$.

%%%%%%%%%%%%%%%
\section{The \texttt{fmtidal} model}
\label{sec:tides}
%%%%%%%%%%%%%%%
We focus on the tidal excitation of a NS's fundamental or $f$-modes of quadrupolar ($\ell=2$) and octopolar ($\ell=3$) order in a spherical-harmonic decomposition for nonspinning stars. We treat the NS itself as fully relativistic but compute tidal interactions in the binary in Newtonian gravity. We emphasize that for the nonspinning binaries considered here the relativistic corrections from the redshift and frame-dragging effects nearly cancel~\cite{Steinhoff:2016rfi}, making the Newtonian tidal excitation of $f$-modes considered here a reasonable approximation. For each $\ell$-th multipole there are, in general, $(2\ell+1)$ $m$-modes that become resonantly excited when $|m| \omega_\mathrm{orb} = \omega_\ell $, where $\omega_{\rm orb}$ is the binary's orbital frequency. 

To compute the dynamical $f$-mode effects in the GW phasing, we follow Ref.~\cite{Flanagan:2007ix}: From the effective action describing a binary system comprising two  finite-sized objects on circular orbits, we compute equilibrium solutions for the $f$-mode degrees of freedom driven below their resonance frequency, calculate the energy of the system, and compute the power radiated in GWs from the quadrupole formula. Using the stationary phase approximation (SPA) the frequency domain GW signal can be written as~\cite{Finn:1992xs,Cutler:1994ys} $h(f)={\cal A}(f) {\rm exp}{[i (\Psi^{\rm pp}+\Psi^{\rm T})]}$, where ${\cal A}$ is the amplitude, $\Psi^{\rm pp}$ the phasing for point masses, and $\Psi^{\rm T}$ the tidal phase contribution. The leading-order $f$-mode tidal phase correction is given in terms of the definite integral in Eq.~(9) of Ref.~\cite{Flanagan:2007ix} for the quadrupolar $f_2$-modes; for the octopolar $f_3$-modes we obtain
\begin{widetext}
\begin{equation}
\label{eq:octint}
\Psi^{\rm T}_\mathrm{oct} = \frac{75m_B \lambda_{3A} }{8\mu M^6 m_A}\int \frac{u^5(-20 + 375 x_3^2 - 2901 x_3^4 + 9287 x_3^6 - 8343 x_3^8 + 2754 x_3^{10})}{( x_3-1)^3 ( x_3+1)^3 ( 3 x_3-1)^3 ( 3 x_3+1)^3}(v^3 - u^3) du \quad+ \quad A \leftrightarrow B.
\end{equation}
\end{widetext}
Here, the labels $A,B$ denote the two bodies, $x_3(u) := u^3/(M \omega_{3A})$, $M=m_A+m_B$ is the total mass, $\mu=m_A m_B/M$ the reduced mass, and $v=(M\omega_{\rm orb})^{1/3}$. 

The integrals in Eq.~\eqref{eq:octint} and Eq.~(9) of Ref.~\cite{Flanagan:2007ix} diverge at the resonances, with the lowest-frequency resonance occurring when $\omega_\mathrm{orb} = \omega_\ell/\ell$.
However, this divergence is only an artifact of using preresonance solutions for the dynamical multipole moments when deriving the expressions for the GW phase, and it can be avoided by accounting for the GW-driven evolution through the resonance as in~\cite{Hinderer:2016eia, Steinhoff:2016rfi}. The TEOB description of the near-resonance effects based on multiscale approximations involves Fresnel integrals that are nonlocal in time complicating the GW computation, while the simpler option of performing a low-order Taylor series expansion near the resonance fails to capture the detailed behavior~\cite{Hinderer:2009ca,Steinhoff:2016rfi}.
  
Here, we instead apply the Pad{\'e} approximation~\cite{Pade:1892aa, Frobenius:1881} around the adiabatic limit $x_\ell=0$ to model the dominant dynamical $f$-mode tidal effects in the GW phase. Pad{\'e} approximations are commonly used to improve the divergent behavior of Taylor expansions and to derive more robust post-Newtonian (PN) waveform approximants (see e.g.~\cite{Damour:2000zb, Damour:2000gg}). To determine the most suitable order of the Pad{\'e} approximant we compare to the GW phase from the TEOB model. We find that for the $f_2$-mode the $(2,2)$-Pad{\'e} approximant of the integrand provides an accurate yet simple approximation for a range of EoS and mass ratios, and for the $f_3$-integral the $(3,1)$-Pad\'e approximation yields the best agreement with TEOB. This leads to  explicit results for the finite $f$-mode-frequency effects comprising the \fm\ model given by
\begin{widetext}
\begin{align}
\label{eq:phase}
  \Psi^{\rm T}_\mathrm{fm}&=-\underbrace{\frac{\left(10 \sqrt{3} \pi -27-30 \log (2)\right)}{96 \eta} \frac{\Lambda_{2A} X_A^6}{( \Omega_{2A})^2}\left(155-147 X_A \right)v^{11}}_{\text{Newt. $f_2$-mode}}
- \underbrace{\frac{1875 (5 - 6\log{(2)})}{16 } \frac{\Lambda_{3A}X_A^7}{(\Omega_{3A})^2} v^{15}}_{\text{Newt. $f_3$-mode}},
\end{align}
\end{widetext}
where $\Lambda_{\ell A,B} = \lambda_{\ell A,B}/m_{A,B}^{2\ell+1}$, $\Omega_{\ell A,B}=m_{A,B} \omega_{\ell A,B}$, $X_{A,B}=m_{A,B}/M$, and $\eta=X_A X_B$. The \fm\,model for the $\Omega_\ell$-dependent contribution can readily be added to any frequency-domain adiabatic tidal model to complete the description of the $f$-mode effects.  

%%%%%%%%%%%%%%%%%%%%%%
\section{Model accuracy}
%%%%%%%%%%%%%%%%%%%%%%
First, we benchmark the \fm\, model against results from the TEOB model~\cite{Barausse:2009aa, Barausse:2009xi, Barausse:2011ys,Taracchini:2013rva,Taracchini:2012ig,Bohe:2016gbl,Hinderer:2016eia}. Since the TEOB model is a time-domain model, we explore multiple avenues to obtain the frequency-domain phase: 
(i) computing the Fourier phase via the fast Fourier transform (FFT) of the time-domain TEOB waveform obtained from the publicly available LIGO Algorithms Library (LAL)~\cite{LAL}; 
(ii) numerically solving the differential equation for the SPA phase $
d^2\Psi/(d\omega_\mathrm{orb}^2)=2(dE/d\omega_\mathrm{orb})/(dE/dt)$, where we express the right-hand side as $2/(d^2\phi_{\rm orb}(t)/dt^2)$ and use the numerical results for $\phi_{\rm orb}(t)$ and $t(\omega_{\rm orb})$ obtained from an EOB inspiral evolution with a \textsc{Mathematica} implementation described in~\cite{Hinderer:2016eia};
(iii) similar to (ii) but with $\omega_\mathrm{orb}$ obtained from the EOB conservative dynamics as in~\cite{Damour:2012yf}; 
(iv) using the general result for the SPA phase~\cite{Finn:1992xs,Cutler:1994ys} $\Psi(f) = 2\pi f t(f) - \phi(f) - \pi/4$,
up to arbitrary constants and functions linear in $f$, where $t(f)$ and $\phi(f)$ are obtained by reinterpolating the results for $\phi(t)$ and $f=\dot\phi/(2\pi M)$ of the GWs from an EOB evolution with the LAL code. In each case we also compute the corresponding result for a point-particle inspiral and subtract it from the full TEOB frequency-domain phase to isolate the tidal contribution. We restrict our comparisons to the inspiral epoch, terminating at the peak in GW amplitude as predicted from a fit based on NR simulations~\cite{Bernuzzi:2015rla}.

To compare two phase models, $\Psi_1$ and $\Psi_2$, we fix the residual gauge freedom of an overall time and phase shift parametrized by $(t_0, \phi_0)$ by minimizing the quantity
\begin{equation}
\label{eq:alignment}
I = \int_{f_0}^{f_1} | \Psi_1(f) - \Psi_2(f) - 2\pi f t_0 + \phi_0 |^2 df,
\end{equation}
where the alignment interval is $[f_0, f_1]=[10,50]$ Hz unless stated otherwise. We compare the different Fourier methods for TEOB for several EoS and mass ratios and find that both SPA methods (ii) and (iv) reproduce the FFT phase (i) to high accuracy, with the difference oscillating about zero and maximally $<0.2$ rad towards the very end, which is comparable to the difference we find between two versions of the TEOB point-particle baseline known as ``v2"~\cite{Taracchini:2013rva,Taracchini:2012ig} and ``v4"~\cite{Bohe:2016gbl}. Thus, for all subsequent comparisons we will use the most direct and entirely GW-based method (iv) to compute the Fourier phase of TEOB using the more recent v4 baseline.

\begin{figure}[t]
\begin{center}
\includegraphics[width=\columnwidth]{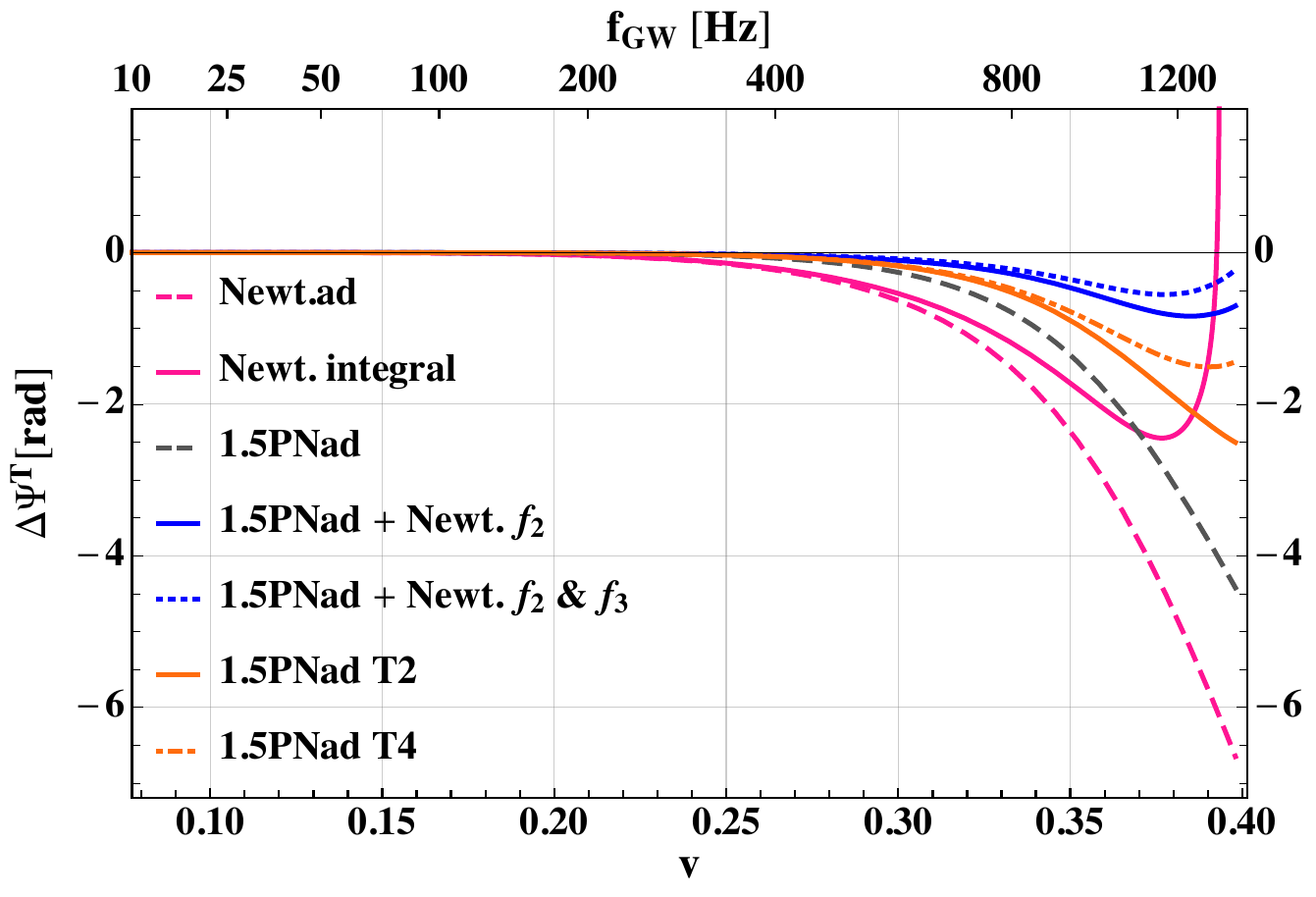}
\caption{Total phase differences to TEOB for several PN tidal approximants for NSs with the H4 EoS and masses $1.8+1.2 {\rm M}_\odot$. The solid magenta curve corresponds to the difference between TEOB and the divergent Newtonian integral solutions, obtained by solving the integrals in Eq.~\eqref{eq:octint} and Eq.~(9) of~\cite{Flanagan:2007ix} numerically}. The dashed curves show the differences to TEOB for the Newtonian (magenta) and 1.5PN adiabatic (grey) tidal phases. Both the Pad{\'e} (blue curves) and the low-order Taylor approximants, which both include $\ell=(2,3)$ (orange curves) for the finite-$f$-mode frequency effects remain regular, with the Pad{\'e} models having significantly smaller discrepancies to the TEOB phase.
\label{fig:test}
\end{center}
\end{figure}

We now consider the impact of different modifications to the PN tidal phase. Figure~\ref{fig:test} shows the phase differences to TEOB and is representative of a number of other cases we analyzed. As seen from the figure, using the adiabatic tidal phase complete at 1.5PN order (1.5PNad) ~\cite{Flanagan:2007ix,Hinderer:2009ca,Damour:2012yf,Vines:2011ud} combined with the $f_2$-mode term of \fm\, yields good agreement with the TEOB model (solid blue curve); including the $f_3$-mode effect leads to a small further reduction in the discrepancy (dotted blue curve). For comparison, we also show the divergent solutions of the integrals from~\cite{Flanagan:2007ix} and Eq.~\eqref{eq:octint} w.r.t. TEOB (solid magenta curve). In addition, Fig.~\ref{fig:test} also shows that, as expected, low-order Taylor expansions [to quadratic (T2) and quartic (T4) order in $x_\ell$] of the apparent resonance singularity yield a larger difference to the TEOB phase (orange curves). 
We also consider higher-order (albeit incomplete) adiabatic PN corrections~\cite{Damour:2012yf} to the total tidal phase and find that the differences to TEOB with the 2.5PN adiabatic baseline are similar to (or smaller than, in some cases) the 1.5PN, and that the incomplete 2PN adiabatic tidal baseline gives the largest disagreement. This trend is not surprising as full PN orders are known to increase tides, while the tail terms present in half-PN orders reduce tidal effects; the TEOB model includes all of these. For the results presented here we choose to use the theoretically complete adiabatic 1.5PN order and note that it is straightforward to change the baseline in this model.

While the Pad{\'e} approximation successfully regularizes the divergent phase integral, we still expect the \fm\, model to deviate from the TEOB phase at frequencies $\gtrsim 1$ kHz\footnote{We note, however, that current generation GW detectors have reduced sensitivity at GW frequencies $\gtrsim 1$ kHz.} as it does not include details of the resonant $f$-mode excitation and relativistic corrections. Since the TEOB model itself is likely to become inaccurate and miss additional physics in this regime, we leave further improvements of our model to future work once new analytical and NR knowledge becomes available.

Next, we test the \fm\,model for a range of binary configurations: three different EoSs of increasing stiffness (APR4~\cite{Akmal:1998cf}, MPA1~\cite{Muther:1987xaa} and H4~\cite{Lackey:2005tk}), and NS masses of $1.375+1.375 {\rm M_\odot}$ and $1.8+1.2 {\rm M_\odot}$. We use the approximate URs to obtain the $f$-mode frequencies~\cite{Chan:2014kua} and octopolar deformability $\lambda_3$~\cite{Yagi:2013sva,Yagi:2014bxa}. To verify that our model correctly captures the dependence on $\lambda_\ell$ and $ \omega_\ell$ we also include a fiducial equal-mass NS-NS case where the URs are explicitly broken -- concretely, we choose $\Lambda_2=700$ and $\Omega_2=0.06$ instead of $\Omega_2^{\rm UR}=0.0696$ as predicted by universal relations -- and a BH-NS binary with masses $2.0+1.0 {\rm M_\odot}$ and the MPA1 EoS for the NS. Unless stated otherwise, we consider the total post-Newtonian tidal phase used in these comparisons to be the sum of the 1.5PN adiabatic tidal phase~\cite{Vines:2011ud,Damour:2012yf}, the Newtonian adiabatic octopolar term~\cite{Hinderer:2009ca}, and $\Psi^{\rm T}_\mathrm{fm}$ from Eq.~\eqref{eq:phase}.

\begin{figure*}
\includegraphics[width=\columnwidth]{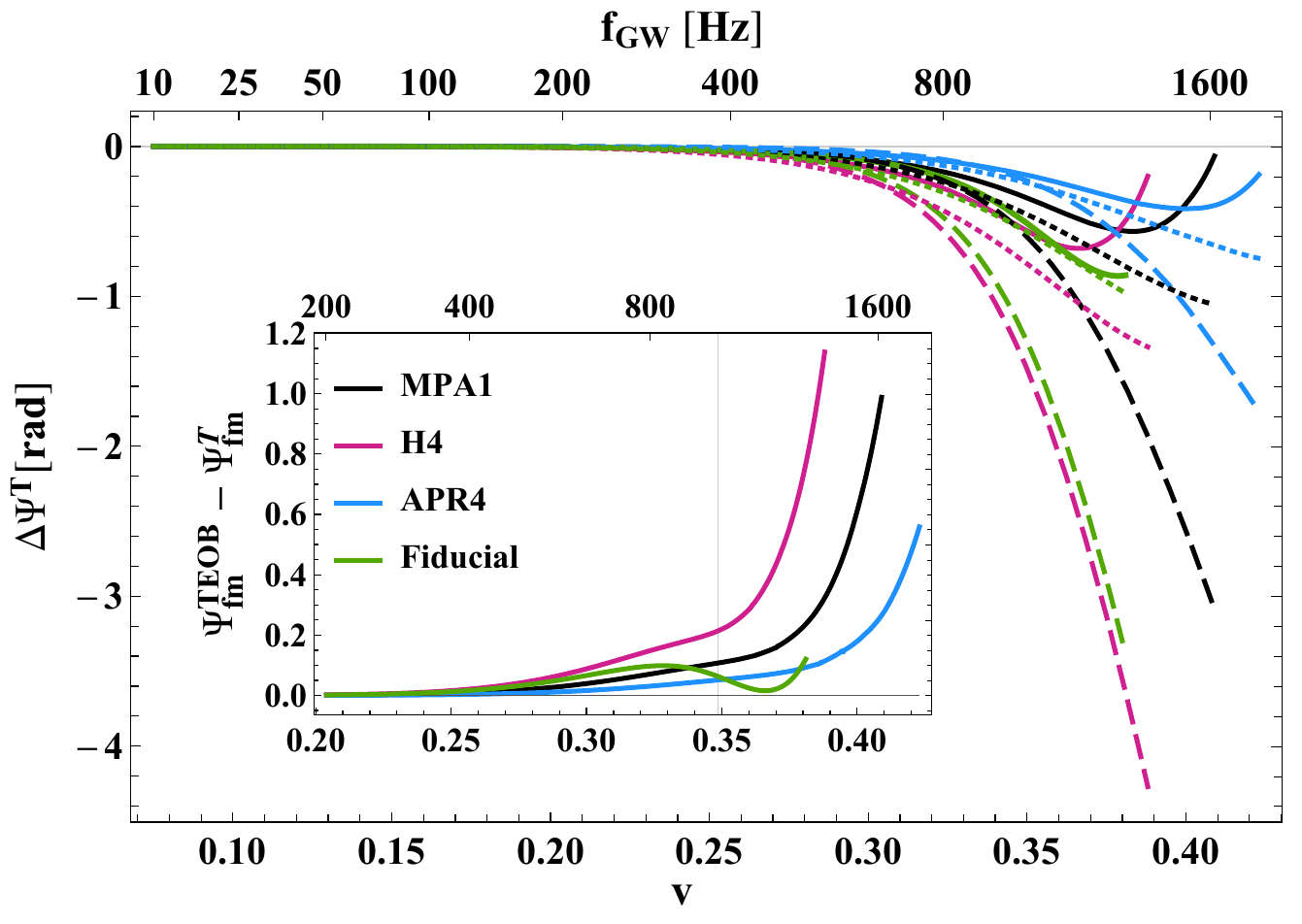} \quad
\includegraphics[width=\columnwidth]{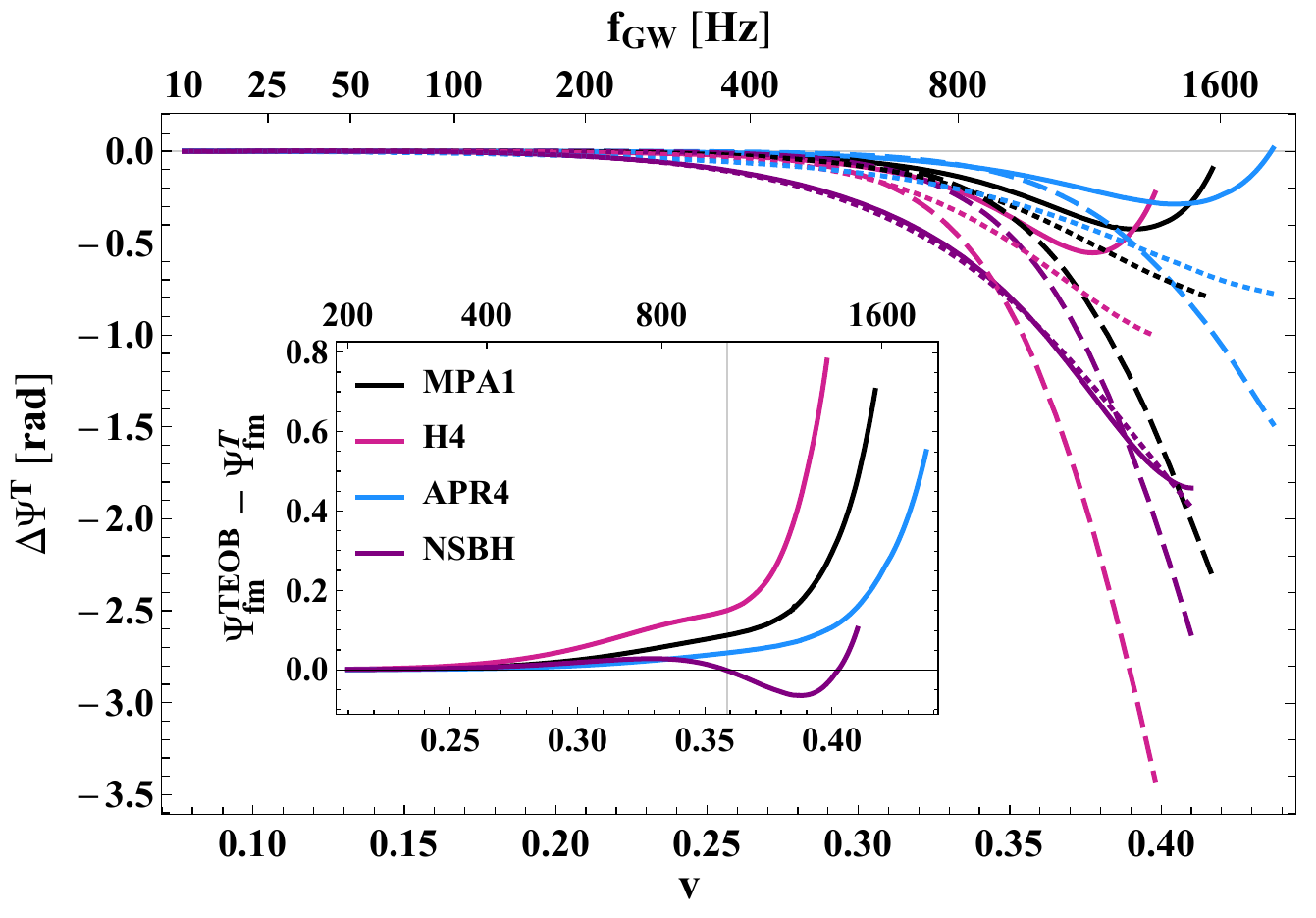}
\caption{Dephasing from the corresponding TEOB predictions. Solid lines show the difference between TEOB and the total PN tidal phase as described in the text. Dashed curves indicate the size of the dynamical tides and correspond to the difference between adTEOB and TEOB. Dotted curves show the error in the 1.5PN adiabatic tidal description (1.5PNad) compared to adTEOB. Inset: Difference between the dynamical tidal phase in TEOB (TEOB minus adTEOB) and the \fm\,model. Left panel: Equal-mass configurations with different EoS and a fiducial case (green curve) where the URs are explicitly violated (see text for details). Right panel: Unequal-mass binaries. The purple curves correspond to a BH-NS binary with masses $2.0+1.0 {\rm M_\odot}$ and the MPA1 EoS for the NS.}
\label{fig:results}
\end{figure*}

The left (right) panel of Fig.~\ref{fig:results} shows the phase differences against TEOB for the equal (unequal) mass configurations. In all cases we find that adding the \fm\,effects to the adiabatic PN phase (solid curves) improves the overall agreement with TEOB as seen from comparing to the dotted curves corresponding to the adiabatic phasing only. The inset in Fig.~\ref{fig:results} is a direct comparison of the $f$-mode effects alone, showing the difference between \fm\,and the purely dynamical phase contributions in TEOB computed by subtracting the adiabatic TEOB phase (adTEOB) obtained with the LAL code by specifying very high values for the $f$-mode frequency. The \fm\,results generally underestimate the dynamic tides; however, up to a GW frequency of $\sim 1$ kHz, and even higher frequencies for softer EoSs (e.g. APR4, blue curve), \fm\,reproduces the TEOB dynamic tides to within $90\%$ accuracy and better for all configurations considered. The systematic underestimation of dynamic tides in \fm\, w.r.t. TEOB may be attributed to higher-order adiabatic tidal corrections incorporated in the EOB model that mix nonlinearly with the $f$-mode effects. We stress that in all cases the dominant total phase difference comes from the adiabatic sector and not the dynamical tides model as evident from the dot-dashed curves in the main panel. 

\begin{figure*}
\includegraphics[width=\columnwidth]{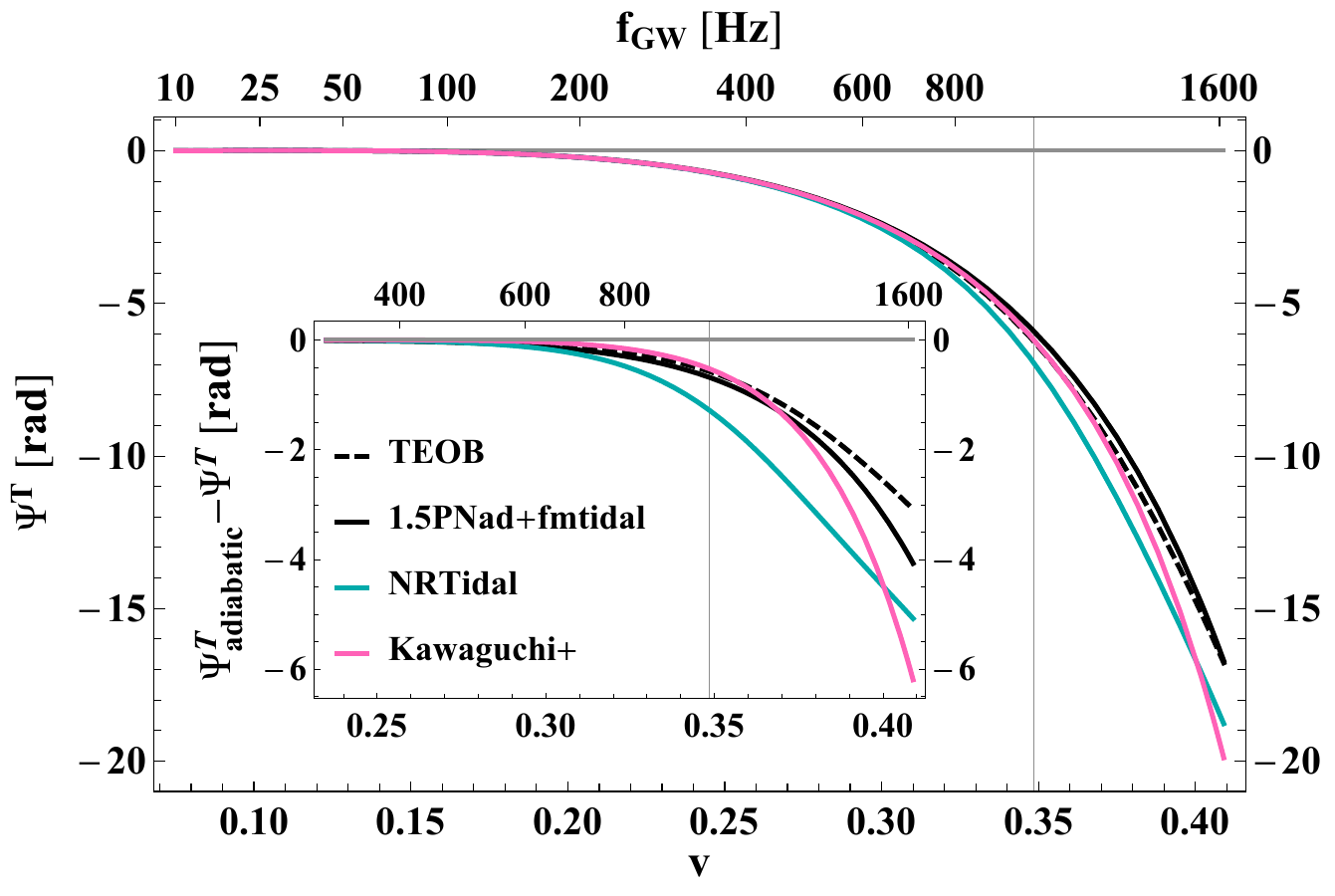}\quad
\includegraphics[width=\columnwidth]{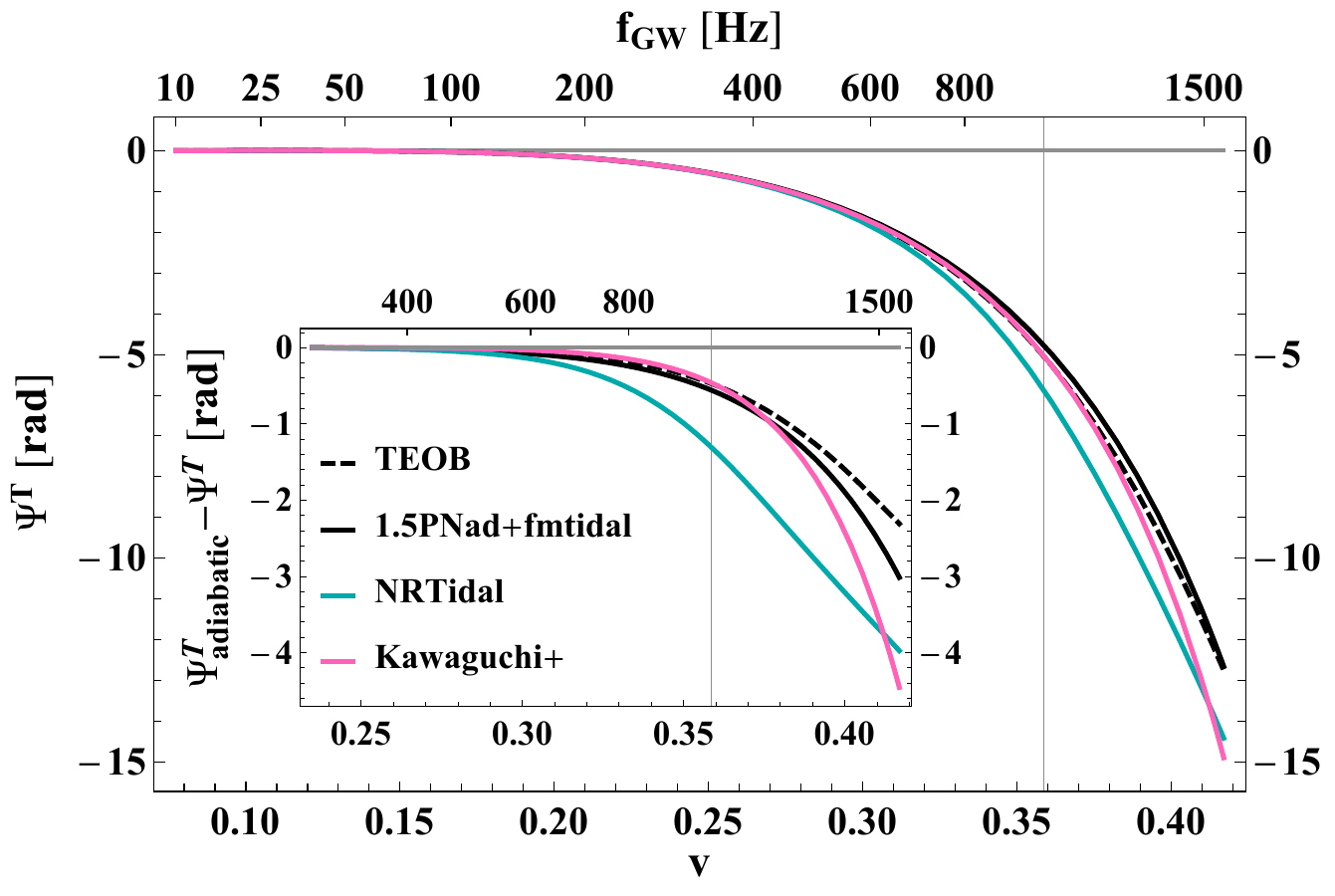}
\caption{Comparison to NR-calibrated models~\cite{Dietrich:2017aum,Kawaguchi:2018gvj} for the MPA1 EoS for equal-mass (left panel) and unequal-mass (right panel) binary neutron star configurations. The main plots show the total tidal phase for each of the four models, i.e. TEOB, 1.5PNad+\fm, NRTidal and Kawaguchi+, while the inset shows the phase residual obtained by subtracting adTEOB from the NR models and TEOB itself; we also show $\Psi^{\rm T}_{\rm fm}$ in the inset (black solid curves). 
Focusing on the insets in both panels we see that the phase residual of the phenomenological Kawaguchi+ model (magenta curves) is very close to our analytic prediction of the $f$-mode effect $\Psi^{\rm T}_{\rm fm}$ (solid black curves) up to $1$ kHz (inset) where the Kawaguchi+ model is valid, with the (uncontrolled) extrapolation up to merger also shown. The discrepancies with the NRTidal model~\cite{Dietrich:2017aum} (turquoise curves) are due to known shortcomings of the NRTidal model.}
\label{fig:NRcomparison}
\end{figure*}

Finally, Fig.~\ref{fig:NRcomparison} compares the \fm\,model to NR-calibrated models and is representative of results for a wider range of EoSs and mass ratios that we considered. The NR-calibrated phenomenological models~\cite{Dietrich:2017aum,Kawaguchi:2018gvj} provide explicit expressions for the tidal phase $\Psi^{\rm T}$ in terms of the parameters $\{X_A,X_B, \Lambda_{2A},\Lambda_{2B}; v\}$ but do not include an explicit dependence on the $f$-mode frequency. The NRTidal model~\cite{Dietrich:2017aum} is known to have shortcomings for unequal-mass binaries~\cite{Kawaguchi:2018gvj}, and from Fig.~\ref{fig:NRcomparison}, we notice its disagreement with \emph{all} analytic models from GW frequencies of $\sim 600$ Hz. Thus, focusing on comparisons to the Kawaguchi+ model~\cite{Kawaguchi:2018gvj} which is only valid in the regime below $1$ kHz (indicated by the vertical line), we find very good agreement for the total phase as well as the phase residual when subtracting the adiabatic TEOB phase (see inset). This indicates that a large part of the phenomenology extracted from NR simulations in this regime is consistent with the dynamical $f$-mode effect. The advantage of the analytic \fm\, description is its explicit physics content and dependence on characteristic matter parameters, which enables us to construct more robust models and perform new tests of fundamental physics.  

%%%%%%%%%%%%%%%%%%%%%%
\section{Discussion}
%%%%%%%%%%%%%%%%%%%%%%
We have developed an approximate closed-form model of dynamical $f$-mode tidal effects in the frequency-domain phase of GWs from a nonspinning binary inspiral. This model can be directly used in state-of-the-art computationally efficient BH baseline waveforms~\cite{Hannam:2013oca,Khan:2015jqa, Purrer:2015tud, Bohe:2016gbl} together with a frequency-domain adiabatic tidal model~\cite{Vines:2011ud,Damour:2012yf}, which are routinely used in GW observations. 
While the \fm\,model derived here is based on a number of restrictions, it is readily amenable to future improvements using analytical results for other matter signatures and inputs from NR simulations for the late inspiral and beyond.
As we demonstrated, our dynamical tides model \fm\,is in good agreement with the TEOB model, which contains a more detailed description of dynamic $f$-mode tides but is computationally expensive rendering, it inefficient for GW data analysis.
Having efficient models such as \fm\,for data analysis will become especially important as the sensitivity of GW detectors increases in the coming years, and we anticipate detecting tens of NS binary inspirals per year~\cite{GWTC-1}. Further, including more realistic physics in frequency-domain tidal models, such as the $f$-mode dependence derived here, provides a useful baseline for future efforts to reduce systematic uncertainties in upcoming GW measurements.

 The main impact of the \fm\,model is its explicit dependence on the different characteristic matter parameters $\lambda_\ell$ and $\omega_\ell$. This enables new measurements~\cite{Pratten:2019sed} and efficient studies for the science case and design of future GW detectors without the restrictive assumption of the validity of quasiuniversal relations between $\lambda_\ell$ and $\omega_\ell$. Relaxing the UR assumption substantially enriches the scope of science derivable from future GW observations, allows us to gain deeper insights into matter and fundamental forces in unexplored regimes by probing multiple characteristic parameters simultaneously, and enables novel tests for strong-field dynamical gravity in the presence of matter and exotic compact objects that are otherwise impossible.

%%%%%%%%%%%%%%%
\section*{Acknowledgments}
%%%%%%%%%%%%%%%
We thank Geraint Pratten, Alberto Vecchio, Samaya Nissanke, David Nichols, and Tim Dietrich for useful discussions and comments on the manuscript. P. S. acknowledges NWO Veni Grant No. 680-47-460. T. H. acknowledges support from the DeltaITP and NWO Projectruimte Grant No. GW-EM NS.

\bibliography{References}

\end{document}